\documentclass[fleqn,10pt]{wlscirep}
\usepackage[utf8]{inputenc}
\usepackage[T1]{fontenc}

\usepackage{chemformula}
\usepackage{bm}
\newcommand*\Rm{r_\text{m}}

\title{Van der Waals five-body size-energy universality}
\author[1,*]{Petar Stipanovi\'{c}}
\author[1]{Leandra Vranje\v{s} Marki\'{c}}
\author[2]{Jordi Boronat}
\affil[1]{University of Split, Faculty of Science, R. Bo\v{s}kovi\'{c}a 33, HR-21000 Split, Croatia}
\affil[2]{Departament de F\'\i sica, Campus Nord B4-B5, Universitat Polit\`ecnica de Catalunya, E-08034 Barcelona, Spain}
\affil[*]{pero@pmfst.hr}

\begin{abstract}
A universal relationship between scaled size and scaled energy is explored in five-body self-bound quantum systems. 
The ground-state binding energy and structure properties are obtained by means of  the diffusion Monte Carlo method. 
We use pure estimators to eliminate any residual bias in the estimation of the cluster size. 
Strengthening the inter-particle interaction, we extend the exploration from the halo region to classical systems. 
Universal scaled size-scaled energy line, which does not depend on the short-range potential details and binding strength, is found for homogeneous pentamers with interaction potentials decaying at long range predominantly as $r^{-6}$. 
For mixed pentamers, we discuss under which conditions the universal line can approximately describe the size-energy ratio. 
Our data is compatible with generalized Tjon lines, which assume a linear dependence between the binding energy  of the pentamers and the one of tetramers, when both are divided by the trimer energies.  
\end{abstract}

\begin{document}

\flushbottom
\maketitle
\thispagestyle{empty}

\section*{Introduction}
Universality in few-body systems connects physical systems at vastly different 
energy and length scales. It manifests as the independence of system's 
characteristics upon the shape of the interaction potential and the length 
scale. 
The most famous universal phenomenon is the Efimov's prediction~\cite{Efimov} of 
the geometric series of three-body bound state levels which occur when two-body 
state has zero energy, i.e., in the unitary limit. 
Although the first Efimov candidates were expected in nuclear 
physics,~\cite{EfimovNuclei,RMPhalos,riisager} the first signature came from an 
ultracold gas of cesium atoms~\cite{EfimovCs}. 
This was possible due to the ability to control interactions between atoms by a 
magnetic field, thanks to the presence of Feshbach resonances. 
The Efimov effect was very soon observed in other cold atom systems, 
including those with non-identical particles and $(N>3)$-body systems, in which 
a variety of universal bound states linked to the Efimov trimer was 
found.~\cite{EfimovRev,EfimovRev2,EfimovPRL-H3L,EfimovPRL-Cs3Li,Efimov8}. 
A further unexpected van der Waals universality appeared for three atoms interacting with potential $-C_6 r^{-6}$ in the ultracold regime, near Feshbach resonances~\cite{EfimovRev,EfimovRev2,vdWBerninger,vdWWangY,vdWWangJ}. 
The ground-state trimer dissociation scattering length $a_{-}^{(0)}$, which acts 
as a three-body parameter, appeared universally proportional to the van der 
Waals length $l_{\rm vdW}$. 
Wang \textit{et al.}~\cite{vdWWangJ} explained the emergence of an 
effective repulsive three-body barrier, which prevents the three particles from 
getting close together, thus preventing configurations with small hyperradii, 
$\rho>2l_{\rm vdW}$. 
In the limit of zero-range interactions and large scattering lengths, there are evidences~\cite{Hadizadeh2011,Bazak2019,Frederico2019} for scales beyond three-body and the consequent necessity of a four-body scale when particles interact through an attractive contact~\cite{Bazak2019,Frederico2019} or soft core~\cite{GattobigioKievsky2011} pairwise potential.

The Efimov effect was also observed by the Coulomb explosion 
imaging~\cite{Sci} in the experimentally elusive atomic trimer $^4$He$_3$, 
which is weakly-bound under natural conditions. 
Clusters which are even more weakly bound than $^4$He$_3$ present also 
different types of universality. 
Importantly, they are examples of quantum halo states, i.e., systems which 
prefer to be in classically forbidden regions of the space. 
Their large spatial extent makes the details of their interparticle interactions less important, leading to universal properties. 
The search for a universal relation between size and energy, in quantum halo 
states, began in nuclear physics~\cite{EfimovNuclei,RMPhalos,riisager} and 
was later continued in atomic systems. 
The precise knowledge of the interparticle interactions in atomic 
clusters~\cite{book1,book2} made it possible to determine universal ground-state 
size-energy ratios in weakly-bound dimers, trimers, and 
tetramers~\cite{Unihalo,Uni4}. 
The progress in Coulomb explosion imaging enabled measurements of the 
distribution functions in dimers, trimers, and tetramers of Argon and 
Neon~\cite{ExpNeAr}, as well as weakly-bound Helium trimers~\cite{NC,Sci,Sci2} 
and dimer $^4$He$_2$~\cite{He2}. 
A thorough analysis for a large set of pure and mixed weakly-bound atomic 
dimers, trimers, and tetramers showed that universal size-energy scaling 
extends even below the halo area~\cite{Unihalo}, in the so called quasi-halo 
region. Four-body systems with large size were found in Helium and 
Helium-alkali tetramers~\cite{He4,4HeA}, but also in pentamers~\cite{5HeA}. 
It is therefore interesting to explore the existence of a  
universal relation between energy and size of five-body clusters, in a wide 
range extending from weakly-bound quantum halo systems to 
classical ones. 

Close to the unitary regime, Tjon~\cite{Tjon1975,Tjon1981} predicted a linear 
relation between the binding energy of the $\alpha$ particle and the triton, 
which was shown to approximately hold for different nuclear models. 
It was argued that, in the universal regime, a four-body parameter is not 
needed for determining the energy of the four-body cluster~\cite{EfimovRev}. 
The so-called Tjon lines were later investigated in atomic systems close to 
unitarity~\cite{Platter,HannaBlume,Hiyama2012,Bazak2016,Lekala}.  Hanna and 
Blume~\cite{HannaBlume} did not find that the energies of $E_{N+1}$ and $E_{N}$ 
clusters are well described by linear relations.  However, they and 
others~\cite{Hiyama2012,Bazak2016,Lekala} showed an approximate validity 
of generalized Tjon lines connecting linearly the
relative energies,  $E_{N+1}/E_{N-1}$ and $E_{N}/E_{N-1}$.
There are some differences between the predictions of the generalized Tjon lines
slope in previous studies, that occur most likely due to analysis of different 
ranges around the universality limit~\cite{HannaBlume}.  
Additionally, Yan and Blume~\cite{YanBlume} showed that, at 
unitarity, the energies of few-body systems are not fully independent of the 
shape of the two-body short-range potentials. 
However, they found that in the case of van der Waals two-body interactions 
the binding energies at unitarity are approximately given solely in terms 
of the van der Waals length. 
It has not been reported how the relationship between  $E_{N+1}/E_{N-1}$ and  
$E_{N}/E_{N-1}$ changes when moving away from the unitary limit, in direction of 
even more weakly bound states or when approaching the classical limit, or how 
well realistic atomic clusters approach the results obtained 
by model Lennard-Jones systems in these limits. 
Such findings are relevant for a better understanding of the limits of 
universality in Lennard-Jones systems. 

In the present work, we study the energies and sizes of five-body Lennard-Jones 
clusters with the goal of determining the extension of their universality, from 
strongly to extremely weakly bound systems, that can be regarded as quantum 
halo states. 
Besides model systems, we study a range of realistic clusters containing up to 
three different atomic species. 
We rely on the use of quantum Monte Carlo simulations which provide exact
results, within some statistical errorbars. 
We also compare the obtained five-body energies with the energies of four and 
three-body Lennard Jones systems in order to test the 
accuracy of generalized Tjon lines. 

The rest of the paper is organized as follows. Section Methods describes the 
quantum Monte Carlo methods used in our work and introduces the energy and size 
scaling. Section Results discusses first five-body size-energy universality and 
then the obtained Tjon lines. The main conclusions of our work are summarized 
in Section Conclusions.

\section*{Methods}
The ground-state properties, energy $E$ and mean square of inter-particle 
separations $\langle r^2\rangle$, were obtained by solving the Schr\"{o}dinger 
equation 
\begin{equation}\label{eq:SE}
	-\frac{\partial \Psi(\bm{R},\tau)}{\partial \tau} = (H- E_{{\rm r}}) 
	\Psi(\bm{R},\tau) \ ,
\end{equation}
written in imaginary-time $\tau=it/\hbar$, for the Hamiltonian $H$. 
The reference energy $E_{{\rm r}}$ is introduced for numerical convenience. 
The positions of particles in five-body systems are stored in the so-called \textit{walker} 
$\bm{R} \equiv \left(\bm{r}_1,\bm{r}_2,\bm{r}_3,\bm{r}_4,\bm{r}_5\right)$. 
The Schr\"odinger equation is solved stochastically 
utilizing the second-order diffusion Monte Carlo (DMC) method~\cite{DMC2} which, 
within statistical errorbars, leads to the calculation of the exact binding energy $B=-E$, 
when the time-step $\Delta \tau\to 0$, the imaginary time $\tau\to\infty$, and the number of walkers $\to\infty$. 
As usual, importance sampling is introduced in DMC~\cite{DMC2} 
to reduce the variance by multiplying the ground-state wave function by a trial wave function 
optimized using the variational Monte Carlo (VMC) method. 
Estimators which do not commute with the Hamiltonian, 
e.g. $\langle r^2\rangle$, can be biased by the mixed distributions produced by the use of importance sampling. 
In order to completely remove any bias from the trial wave function, we do not use the extrapolation approximation 
$\langle r^2\rangle_\text{ex}\approx 2\langle r^2\rangle_\text{DMC}-\langle r^2\rangle_\text{VMC}$, 
but implement much more sophisticated pure estimators~\cite{pure} to get unbiased estimations. 
Masses and trial wave-functions were taken from our previous works~\cite{Unihalo,HeT,5HeA,Uni4,He4}. 
The use of pure estimators proved to be successful in Helium clusters~\cite{He3}, 
where theoretical predictions on distribution functions reproduced accurately experimental results \cite{NC,Sci} drawn from  Coulomb explosion imaging. 

We are interested in universal relations, so it is not crucial for us to use the most realistic potential, 
but to calculate accurately the ground-state energy and size of a system, for a given potential and particle masses. 
Our potential function sums only pair interactions. 
We take the Lennard-Jones (LJ) 12-6 model 
$V(r)=4\epsilon[(\sigma/r)^{12}-(\sigma/r)^{6}]$, 
with adjustable depth $\epsilon$ and zero-point distance $\sigma$, for a systematic exploration of van der Waals pentamers. 
For real clusters, we use the following model potentials: 
JDW~\cite{JDW} for spin-polarized hydrogen isotopes \ch{^{2,3}H$\downarrow$}, denoting \ch{^3H} also as T; 
Silvera~\cite{Silvera} for hydrogen molecules \ch{H2}; 
TWW~\cite{TWW}, DWW~\cite{DWW}, TY~\cite{TY}, MF~\cite{MF} and MFmod~\cite{MFmod} for \ch{He-H$\downarrow$}; 
semi-empirical HFDB~\cite{HFDB} for helium isotopes \ch{^{3,4}He}; 
KTTY~\cite{KTTY} for interaction of an alkali metal isotope and a helium isotope; 
and TT~\cite{TT} for noble gases Ne and Ar.

To be able to compare quantities differing in several orders of 
magnitude, we scale the energy and the size with a characteristic length and 
analyze dimensionless quantities. 
Similar to what was done in previous works~\cite{RMPhalos,Unihalo}, we measure 
the size of the system by the mean-square hyperradius with subscript $r$,
\begin{equation}\label{hyper1}
	\rho^2_r = \frac{1}{Mm}\sum_{i<k}^{N} m_im_k\langle r_{ik}^2 \rangle \ ,
\end{equation}
with $r_{ik}=|\bm{r}_k-\bm{r}_{i}|$ and $M$  
the total mass of the $N$-body system. The particle masses $m_i$ are given in  
an arbitrary mass unit $m$. 
The characteristic hyperradius $\rho^2_R$ is defined by substituting in Eq. 
(\ref{hyper1}) the pair size $r_{ik}^2$ by the square of the corresponding van 
der Waals length,
\begin{equation}\label{rhoR}
	\quad R_{ik}^2=\sqrt{\frac{2\mu C_6}{\hbar^2}} \ ,
\end{equation}
where $C_6$ is the dispersion coefficient and $\mu=m_im_k/(m_i+m_k)$ the 
reduced mass of a given pair. 
Notice that there is  a different definition in the 
literature~\cite{EfimovRev} for this length, $l_{\rm vdW}=0.5R$. 
In previous research of four-body systems~\cite{Uni4}, we showed that 
the van der Waals length $R$ is convenient for scaling weakly and 
strongly bound systems, and it was also used in the context of universal 
relations~\cite{EfimovRev}. 
We scaled the size of the pentamers as $Y_\rho=\rho^2_r\rho_R^{-2}$ 
and analyze in the next Section how it depends on the dimensionless scaled 
binding  energy, $X_E=mB\rho^2_R\hbar^{-2}$. 

\section*{Results}
\begin{figure*}[h!]
	\centering
	\includegraphics{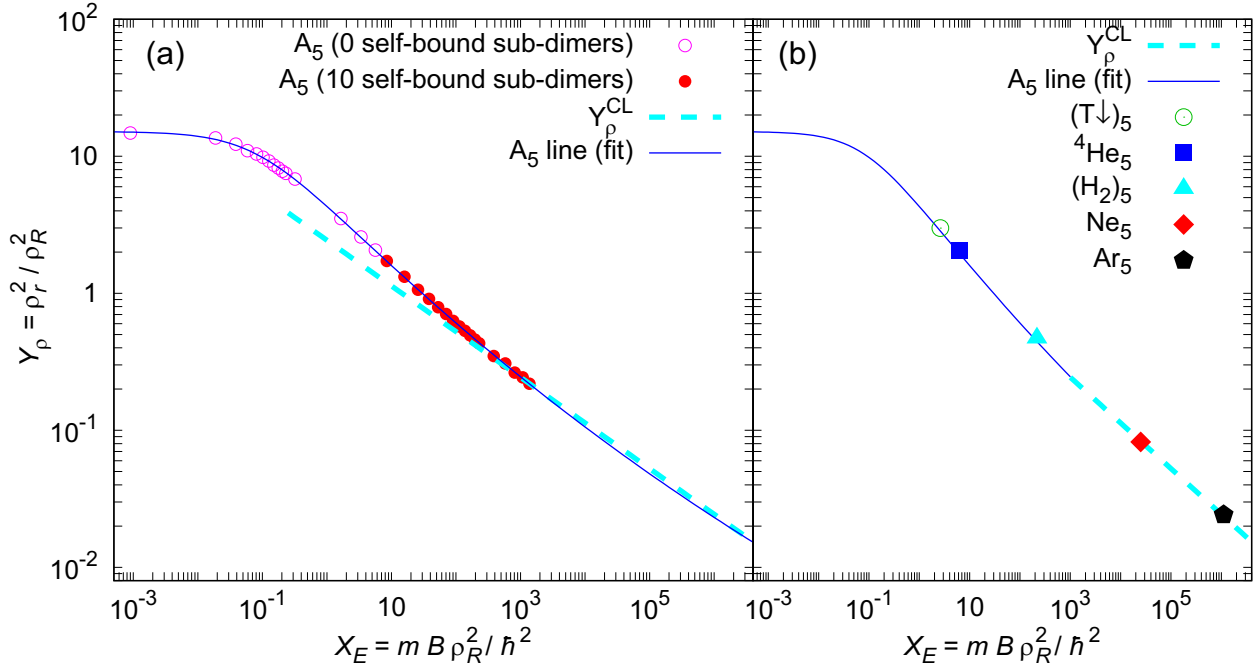}
	\caption{Scaled size-energy fit for various homogeneous quantum five-body systems A$_5$. Interactions are modeled by (a) LJ 12-6 pair potential and (b) potentials for realistic systems. Pentamers are classified according to the number of self-bound sub-dimers. For comparison, we report the classical approximation, with $Y_\rho^\text{\rm CL}$ given by Eq.~\eqref{eq:YCL}.}
	\label{Fig1}
\end{figure*}

\textbf{First}, we discuss \textbf{homogeneous five-body quantum systems} A$_5$, i.e., van der Waals clusters of five identical atoms or molecules A. 
In a previous study of four-body systems~\cite{Uni4}, 
no mass effect on scaling was noticed. 
Therefore, and for practical reasons, we first explored clusters of particles with equal mass
$m_i=4u$, multiple of the atomic mass constant $u$. 
As a pair-potential model, we chose LJ 12-6 
$V(r)=\epsilon[(\Rm/r)^{12}-2(\Rm/r)^{6}]=4\epsilon[(\sigma/r)^{12}-(\sigma/r)^{6}]$, 
where $-\epsilon$ is the minimum at inter-particle separation $r=\Rm=\sqrt[6]{2}\sigma$, and $r=\sigma$ is the zero-point of the potential. 
The dispersion coefficient in this case has the simple form $C_6=2\epsilon\Rm^{6}$. 
We used a repulsive core $\sigma = 4$~\AA\, and varied the potential depth $\epsilon$. 
This allowed us to explore a wide range of binding strengths, from $0.08$~mK for weakly-interacting  ($\epsilon=3.32$~K, $\sigma=4$~\AA) to $50.752$~K for strongly-interacting ($\epsilon=20$~K, $\sigma=4$~\AA) pentamers.
Corresponding $\langle r^2\rangle$ appear in reverse order, from $990$~\AA$^2$ to $36$~\AA$^2$. 
The scaled size $Y_\rho=\rho^2_r\rho_R^{-2}$ and energy $X_E=mB\rho^2_R\hbar^{-2}$ for these model systems are shown with points in panel (a), Fig.~\ref{Fig1}. 
They span many orders of magnitude so a logarithmic scale is used. 

In an homogeneous  pentamer \ch{A5}, there are $10$ equal A-A pairs of particles so we 
distinguish between clusters with 
zero or ten self-bound sub-dimers, represented with empty and full symbols, respectively. 
As one can see, all pentamers in Fig.~\ref{Fig1} follow the same law, regardless of the interaction potential and number of sub-dimers. 
The empirical function, similar to the four-body universality~\cite{Uni4},  
\begin{equation}\label{eq:fit}
	Y(X)=Y_0\exp\left\lbrace X_0\left(1+\xi X^k\right)^{-1/n}\right\rbrace \ ,
\end{equation}
fitted well (thin blue line) the obtained data for the scaled energies below $10^6$. 
The parameters of the best fit are reported in the Table~\ref{tab:fit}. 
\begin{table}[ht]
	\begin{tabular}{ccccc}
		\hline
		$Y_0$  &  $X_0$  &  $\xi$  & $k$ &  $n$ \\
		\hline
		$10^{-6}$  &  $16.5348(64)$  &  $8.86(38)$  &  $0.8830(61)$  &  $28.88(29)$\\
		\hline
	\end{tabular}
	\caption{\label{tab:fit}Parameters of Eq.~\eqref{eq:fit} that fit the DMC data in Fig.~\ref{Fig1}. Figures in parenthesis are the statistical errors.}
\end{table}

In the limit $T\to 0$ and for strong interactions $(B\to\infty)$, a system becomes \textbf{classical} and its structure is defined by its minimum potential energy.  
Two-, three- and four-body classical systems rest in equilateral geometrical arrangements with all inter-particle separations equal to the position of the pair-potential minimum $r=\Rm$. 
Respectively, particles are located at the vertices of the line segment, triangle and tetrahedron, which are one-, two- and three-dimensional geometry objects. 
The structure of a five-body system is more complicated because it is not possible to form a geometrical structure in three-dimensional space where all vertices are equally separated. 
As an optimal structure in this case, we take a triangular dipyramid, i.e., a double tetrahedron with common base. 
Then, nine pairs span 9 edges of length $\Rm$ and contribute to the binding energy by $\epsilon$. 
The remaining pair spans the only spatial diagonal whose length corresponds to double height of tetrahedron, $2H=\sqrt{8/3}\Rm$, and thus contributes to the binding by an amount $\epsilon\left|\left(\frac{\Rm}{2H}\right)^{12} -2\left(\frac{\Rm}{2H}\right)^{6} \right|$. 
If we take the mass of a particle as the mass unit, the hyperradius simplifies,
\begin{equation}\label{eq:CLro}
	\rho_r^2=\frac{m^2}{Nm^2}\left[9\Rm^2+(2H)^2\right]=\frac{7}{3}\Rm^2.
\end{equation}
Scaling the size \eqref{eq:CLro} with the characteristic hyperradius  
$ \rho_R^2=2R^2$, as well as the binding energy,
\begin{equation}\label{eq:CLB}
	X_E=\frac{mB\rho^2_R}{\hbar^{2}}=\frac{2m\epsilon}{\hbar^2}\cdot\frac{2386215}{262144}\cdot R^2,
\end{equation}
leads to a straightforward relationship between scaled size and energy of classical systems, 
\begin{equation}\label{eq:YCL}
	Y_\rho^{\rm CL} \approx \left( 14.5/{X_E}\right)^\frac{1}{3} \ . 
\end{equation}
This classical line is plotted in Fig.~\ref{Fig1} as a thick dashed cyan line which, for scaled energies larger than $10^3$, smoothly continues the trend shown by quantum pentamers. 
All data of analyzed homogeneous five-body systems follow the same line. 
Thus, the universal law applies starting from purely quantum systems, defined by the relation \eqref{eq:fit}, and then extends to classical systems, where for $X_E>10^3$ it asymptotically takes a much simpler form \eqref{eq:YCL}. 
The universal quantum law starts differing from the simple classical estimation for scaled energies $X_E<10^3$, when the contribution of the kinetic energy becomes significant, producing larger spatial structures than classical ones. 

When mean particle separations become few times larger than van der Waals radius $R\sim \epsilon^{1/4}$, while decreasing $\epsilon$, the binding energy rapidly vanishes ($B\to 0$), but the size barely changes. In this case, particles are far away and pair potentials barely affect the probability outside the range of the van der Waals potential. That scenario is similar to the one of finite and contact interactions and thus, it is in agreement with previous theoretical findings~\cite{Efimov8,Brunnian} that scaled size saturates in the weak binding limit. Weak binding, which does not support smaller clusters, holds pentamers through mediated interactions of additional particles.

If we change the short-range part, i.e., reduce the core size two times, $\sigma=2$~\AA, we can see no effect in the scaling law thus confirming the universal ratio. Pentamers $(8u)_5$ for potential depths $\epsilon=8, 9, 14$~K, respectively, have ground-state binding energies $B=554,1318,8768$~mK and sizes $\langle r^2\rangle=47.6,34.0,17.1$~\AA$^2$, which when scaled, $X_E=3.36,8.47,70.3$, $Y_\rho=2.59, 1.74,0.70$, fit to the universal line.

In addition, we test the validity of the obtained law for the case of \textbf{realistic homogeneous pentamers}, whose interaction is formulated with elaborated potentials describing particles as induced fluctuating electric multipoles. 
Although their pair-potentials have different sort-range parts, they share the common feature that fall quickly with separation $r$ and so the London dispersion energy $-C_6r^{-6}$ dominates at large $r$. 
The ground-state binding energy and size for the studied realistic systems \ch{(T$\downarrow$)5}, \ch{^4He5}, \ch{(H2)5}, \ch{Ne5} and \ch{Ar5} are reported in Table~\ref{tab:ErHomo} %
\begin{table}[ht]
	\begin{tabular}{ccccc}
		\hline
		Cluster  & $B$ / K  &  $\langle r^2\rangle$ / \AA$^2$ & \textbf{$X_E$} &  $Y_\rho$\\
		\hline
		\ch{(T$\downarrow$)5}   & 0.399(9)  & 158.5(9) & 2.63 & 2.99 \\
		\ch{^4He5} & 1.335(1) &  59.4(4) & 6.37 & 2.05 \\
		\ch{(H2)5}  & 44.34(2) &  27.8(4) & 218 & 0.470 \\
		\ch{Ne5} & 224.29(2) & 11.1(2) & 24970 & 0.082 \\
		\ch{Ar5} & 1110.1(3) & 14.67(2) & 1108500 & 0.024\\
		\hline
	\end{tabular}
	\caption{\label{tab:ErHomo}The ground-state binding energy $B$, mean square pair size $\langle r^2\rangle$, scaled energy $X_E$, and scaled size $Y_\rho$ for five-body realistic clusters. Figures in parenthesis are errorbars.}
\end{table}
and compared with the universal line in  panel (b) of Fig.~\ref{Fig1}. 
They follow the universal line equally well, both in the regime of weak  and strong binding.

\begin{figure*}[h!]
	\centering
	\includegraphics{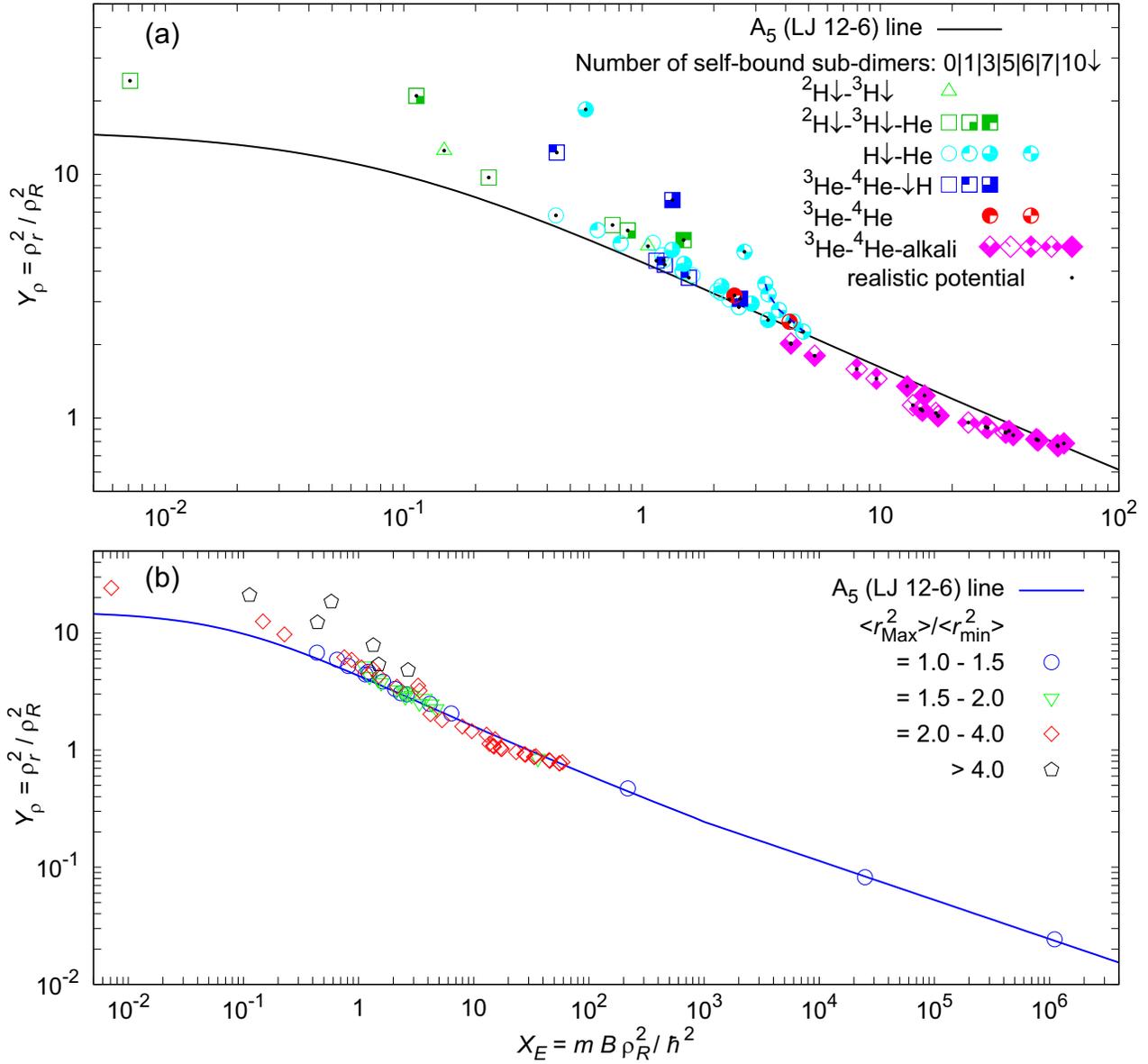}
	\caption{ \label{Fig2} Scaled energies $X_E$ and sizes $Y_\rho$ of various pentamers. (\textbf{a}) The fit obtained in Fig.~\ref{Fig1} for homogeneous systems A$_5$ is compared with five-body mixed realistic systems, using accurate (additional black dot) and previous models of interaction. Blue dashed line is a guide to the eye connecting pentamers \ch{^4He4T$\downarrow$} which separate due to weakening of the \ch{He-T$\downarrow$} interaction. (\textbf{b}) Separation of mixed clusters from the homogeneous universal line is analyzed, comparing ratios of pair sizes.}
\end{figure*}

\textbf{Second}, we test the validity of the obtained universal law for
\textbf{mixed realistic five-body systems} consisting of up to three different 
components: spin-polarized H and He isotopes, an alkali atom, Ne, Ar, and 
hydrogen molecules H$_2$. 
Our results are summarized in Fig.~\ref{Fig2}.

In panel (a), different symbols are used to distinguish different species 
of particles that form the pentamer, while different filling of symbols is used 
to distinguish types of pentamers with regard to the number of self-bound 
sub-dimers. 
Among studied clusters, we distinguish pentamers that have 0 (empty symbols), 1 
(quarter full), 3 (three quarter full), 5 (square with dash), 6 (two quarter 
full), 7 (two quarter empty), and 10 (full symbols) self-bound sub-dimers, as 
it is noted in the legend. 
The results with the realistic potential models are noted with 
additional black dots.
Different symbols are spread everywhere and, at the first sight, it seems that 
no rule can be extracted regarding components, pentamer types or other 
characteristics. 
Only pentamers with all self-bound sub-dimers are always close to the \ch{A5} line, while all other types can be on the line or above it. 
Only mixtures of Helium isotopes and an alkali atom sometimes go below the 
line. 
The latter can be understood in the following way. 
An alkali atom has a much larger electronic cloud repulsive core than Helium 
isotopes, so Helium isotopes tend to form a cluster on one side of an alkali 
atom~\cite{5HeA}. 
This feature limits the arrangements that the pentamer can exhibit, and its 
size is reduced. 
Other mixed systems separate from the line at different points; separation 
points also differ for the same type of pentamers. 

Fragmented-like systems separated from the line share a common feature; all of them have at least one particle which appears less bound than the others and that is significantly separated from the others. 
In panel (b) of Fig.~\ref{Fig2} we made a different analysis.
In this case, the ratio of largest $\langle r^2_\text{Max}\rangle$ and smallest $\langle r^2_\text{min}\rangle$ pairs are compared. 
One can notice that if all pairs are similarly bound, strongly or weakly, i.e., 
if the ratio is between $1$ and $1.5$, mixed systems (circles) follow the 
\ch{A5} line. 
Triangles and diamonds are also close to the line. Thus, if the ratio is below 
$4$, noticeable deviations from the \ch{A5} line can happen only for very weakly 
bound quantum systems, i.e., in the area where scaled energies are less than 
$X_E < 1$, while in other areas only small deviations can occur. 
According to the position on the graph, each system can be recognized from the panel (b) of Fig.~\ref{Fig1} and the panel (a) of Fig.~\ref{Fig2}. 
The larger the ratio is, the larger are the energies for which begining of the start of separation can be expected.
This happens because a very weakly bound component makes negligible 
contribution to the system energy, but significantly increases its size. 
In this case, a small displacement along $X_E$ axis results in a significant 
displacement along $Y_\rho$ axis and the separation occurs. 
Thus, separations which occur for large scaled energies diverge faster, 
feature which was also noticed in the case of tetramers~\cite{Uni4}. 

To illustrate the separation from the universal line, some estimated quantities 
are extracted in Table~\ref{tab:4He4T} for the cluster \ch{^4He4T$\downarrow$}, using different potential models for the \ch{^4He-T$\downarrow$} interaction. 
This allowed modeling different strengths of the \ch{^4He-T$\downarrow$} 
interaction, which is in neither case strong enough to support a dimer bound 
state. 
Thus, \ch{^4He4T$\downarrow$} has $6$ self-bound sub-dimers. 
The results from Table~\ref{tab:4He4T} are shown by five two-quarter full cyan 
symbols that are furthest to the right and above the line in the panel (a) of 
Fig.~\ref{Fig2}, connected by a short dashed blue line to guide the eye.
They deviate very fast from the universal line when the \ch{^4He-T$\downarrow$} 
attraction decreases, starting from the symbol with black point on the line.
In the case of the strongest \ch{^4He-T$\downarrow$} interaction model MFmod~\cite{MFmod}, which is the most realistic one, mean square \ch{^4He-T$\downarrow$} pair size is already $1.58$ times larger than \ch{^4He-^4He} and it is exactly on the universal line. 
Using less attractive \ch{^4He-T$\downarrow$} models, respectively, 
MF~\cite{MFmod}, DWW~\cite{DWW}, TY~\cite{TY}, TWW~\cite{TWW}, binding weakens 
up to $25\%$, almost reaching  the pentamer threshold limit, i.e., 
the ground-state energy of \ch{^4He4} $-577.6(3)$~mK~\cite{He4}, while ratio of 
squared pair sizes doubles. 
Further weakening of the \ch{^4He-T$\downarrow$} interaction would cause 
distancing of the T atom from the remaining tetramer, i.e., scaled size would 
diverge fast in logarithmic scale because scaled energy $X_E=3.29$ is already 
close to the threshold limit $X_E=2.74$, when the pentamer dissociates into 
\ch{^4He4} and far away free T atom. 
The ground-state properties for all studied realistic systems are given in 
Supplementary Table S1 online. 
\begin{table*}[ht]
	\begin{tabular}{|ll|cc|cc|c|c|c|c|}
		\cline{1-7}
		\multicolumn{2}{|c}{Potential model} 
		& \multicolumn{2}{|c}{$R$ / \AA} 
		& \multicolumn{2}{|c}{$\langle r^2\rangle$ / \AA$^2$}
		& \multicolumn{1}{|c}{$\langle r^2\rangle_\text{Max}/$} 
		& \multicolumn{1}{|c}{ } 
		& \multicolumn{1}{c}{ } 
		& \multicolumn{1}{c}{ } \\
		\cline{1-6}\cline{8-10}
		He-He & He-T & He-He & He-T & He-He & He-T & 
		$\langle r^2\rangle_\text{min}$ & $|E|$ / mK & X & Y\\
		\hline
		HFDB~\cite{HFDB} & MFmod~\cite{MFmod} & 5.38 & 6.69 & 65(1) & 103(2) & 1.58 & 885.7(7) & 4.74 & 2.27\\
		HFDB~\cite{HFDB} & MF~\cite{MF} & 5.38 & 6.10 & 65(1) & 106(1) & 1.63 & 866.9(8) & 4.30 & 2.48\\
		HFDB~\cite{HFDB} & DWW~\cite{DWW} & 5.38 & 5.75 & 66(1) & 121(2) & 1.83 & 791.4(9) & 3.75 & 2.79\\
		HFDB~\cite{HFDB} & TY~\cite{TY} & 5.38 & 6.10 & 67(1) & 172(3) & 2.57 & 682.8(9) & 3.39 & 3.22\\
		HFDB~\cite{HFDB} & TWW~\cite{TWW} & 5.38 & 6.10 & 67(2) & 202(3) & 3.01 & 662.8(9) & 3.29 & 3.55\\
		\hline
	\end{tabular}
	\caption{\label{tab:4He4T} Van der Waals length $R$, mean square pair size $\langle r^2\rangle$, ground-state energy $E$, scaled energy $X$ and scaled size $Y$, in the pentamer \ch{^4He4T$\downarrow$} modeled with different pair potentials.}
\end{table*}

Among the studied realistic clusters, the largest ratios of mean square radii 
are in \ch{^4He3(D$\downarrow$)2}, where $\langle r^2 \rangle=106, 940, 1900$~\AA$^2$, 
respectively for pairs \ch{^4He-^4He}, \ch{^4He-D$\downarrow$} and \ch{D$\downarrow$-D$\downarrow$}. 
Thus $\langle r^2_\text{Max}\rangle/\langle r^2_\text{min}\rangle=17.9$, 
while the binding energy is only 7~\% larger than the energy of the trimer~\cite{He3} \ch{^4He3}. 
On average He atoms are close to each other forming a sub-trimer \ch{^4He3} 
which is surrounded with a halo cloud of far away  \ch{D$\downarrow$} atoms. 
Their weak binding is barely mediated by \ch{^4He3} so they very rarely 
find themselves on the same side of the \ch{^4He3}, contributing largely to the size of the cluster. 
Deviation from the line is obvious, $X_E=0.58$, $Y_\rho=18.5$. 
After substituting \ch{D$\downarrow$} with the heavier isotope \ch{T$\downarrow$}, 
which has lower kinetic energy, the cluster becomes homogeneous-like 
with almost ten times lower ratio 
$\langle r^2_\text{Max}\rangle/\langle r^2_\text{min}\rangle=1.85$.

\begin{figure}[h!]
	\centering
	\includegraphics{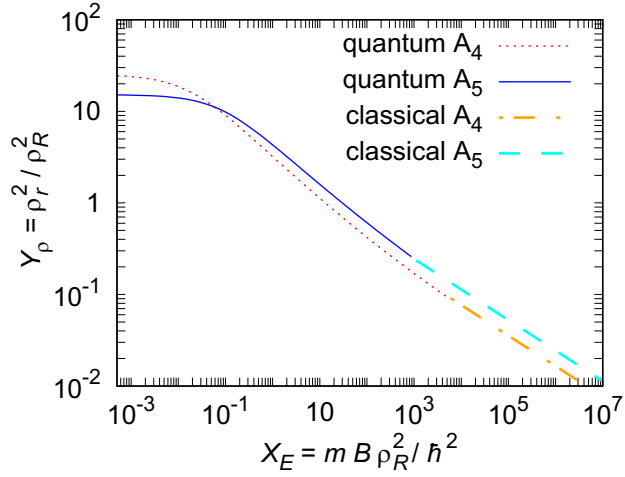}
	\caption{Comparison of the universal size-energy scaling laws for 
		four-~\cite{Uni4} and five body systems.}
	\label{Fig3}
\end{figure}

\textbf{Third}, we compare the present results  with  
previous findings. The size-energy scaling laws for five-body systems 
\eqref{eq:fit} and \eqref{eq:YCL} are compared in Fig.~\ref{Fig3} %
with the four-body universal law~\cite{Uni4}. 
Both lines intersect at $(0.059,11.1)$. In the classical limit, for the same 
scaled energy, strongly bound pentamers have larger scaled size than tetramers. 
Weakening the binding, the difference in scaled size decreases and the inverse 
occurs for $X_E<0.059$. 
In the limit of the binding threshold, the scaled size of pentamer converges to 
$Y_\rho=15$. 

Rasmussen \textit{et al.}~\cite{Efimov8} studied how the two lowest-lying 
weakly bound states of few bosons depend on the strength of two-body Gaussian 
interactions $V(r)=V_0\exp(-r^2b^{-2})$, where $b$ was chosen as the 
characteristic length scale. 
Their results are in qualitative agreement with ours. 
They also predicted that the pentamer has lower scaled size than the tetramer 
at the binding threshold. 
The quantitative comparison of the system size with our results is not 
possible, because we used long-range decay $-C_6r^{-6}$, which is characteristic 
for London dispersion forces between atoms and molecules that are electrically 
symmetric. 
Although they explored only the weak-binding regime, crossing of tetramer and pentamer curves is also just noticeable close to the end of their researched area.
Brunnian systems also show qualitatively the same behavior~\cite{Brunnian}.

\begin{figure*}[h!]
	\centering
	\includegraphics{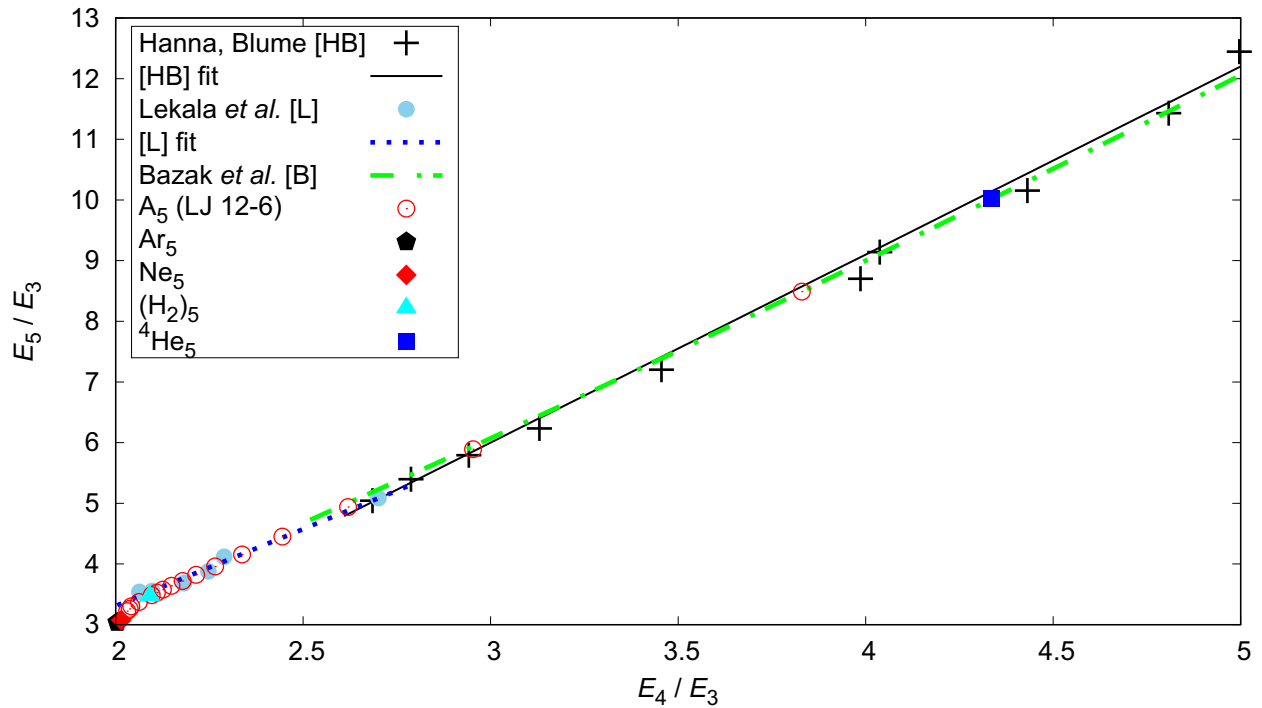}
	\caption{Generalized Tjon line. Our results for clusters in Fig.~\ref{Fig1} are compared with calculations of Hanna and Blume [HB]~\cite{HannaBlume}, Bazak \textit{et al.} [B]~\cite{Bazak2016}, and Lekala \textit{et al.} [L]~\cite{Lekala}.}
	\label{Fig4}
\end{figure*}

Having previously studied also the trimer~\cite{Unihalo} and 
tetramers~\cite{Uni4}, we are able to analyze their energies in comparison with 
the present pentamer results.
Fig.~\ref{Fig4} %
reports the \textbf{generalized Tjon line}, which describes the dependence of the energy ratios $E_5/E_3$ and $E_4/E_3$, where $E_N$ is the ground-state binding 
energy of the $N$-body system. 

Hanna and Blume~\cite{HannaBlume} explored a regime close to the unitary limit, 
i.e., a range $2.6<E_4/E_3<5$ (black symbol $+$), and found 
a linear dependence, with slope $3.10(8)$ (black line). They also noticed 
that the slope decreases if systems very close to threshold are excluded. 
The latter was reconfirmed by the calculations of Lekala \textit{et 
	al.}~\cite{Lekala} 
(sky-blue full circles), whose fit in the range $2.06<E_4/E_3<2.71$ returned 
a slope of $2.5346$ (dotted blue line). 
Bazak \textit{et al.}~\cite{Bazak2016} found that their results, even though not 
quite close to unitarity, follow the empirical relation~\cite{Gattobigio} 
$E_5/E_3\approx\left[2\sqrt{E_4/E_3}-1\right]^2$ (dot-dashed green line). 
Our results, obtained with the model \ch{A5} using the LJ 12-6 potential (empty 
symbols) and realistic models \ch{^4He5}, \ch{(H2)5} are in agreement with 
their findings. 
The realistic \ch{He-He} pair interaction is very close to the unitary limit 
and perfectly agrees with the empirical relation. 
Approaching the trimer threshold (not shown in Fig.~\ref{Fig4} to avoid loss 
of clarity), the estimated ratios $E_4/E_3=11.85$ and $E_5/E_3=34.63$ for 
a model system, with $\sigma=4$~\AA\ and $\epsilon=4$~K, when $E_3=-8$~mK, also 
verify the empirical relation. 

Recent estimates~\cite{Bazak2019} of $^4$He$_N$ binding energies, obtained within the framework of effective field theory at leading order and next-to-leading-order with a four-body force, that renormalizes the four-body system, respectively, $E_4/E_3=4.8(1), 4.35$ and $E_5/E_3=10.8(5), 11.3(3)$ deviate from the empirical relation, but also have large extrapolation errors. Our results with the HFDB potential are $E_4/E_3=4.335(6)$ and $E_5/E_3=10.02(2)$.

Our results show that the linear law is valid 
only for a limited range of ratios $E_4/E_3$. 
Increasing the attraction strength, and leaving the regime of weak binding, the 
slope collapses non linearly and abruptly towards the classical boundary, where 
$E_4/E_3=(6\epsilon)/(3\epsilon)=2$ and $E_5/E_3=795405/262144\approx 3.03$. 
Our realistic clusters \ch{Ne5} $(2.015,3.07)$ and \ch{Ar5} $(2.004,3.04)$ are very close to the classical ratio limit. 
This is to be expected, as the binding energy $1.110$~kK of \ch{Ar5} is very close to the classical limit $1.306$~kK.

\section*{Discussion}

Five-body systems composed of one, two, and three different particles were 
explored by means of quantum Monte Carlo methods at  $T=0$~K. 
Different strengths were analyzed, from very weak binding in quantum 
systems close to the threshold limit, in the halo region, up to the limit of 
maximum binding of purely classical clusters. 
The interparticle interactions were modeled by pair potentials with 
different short-range shape, but with the common feature 
of a long-range behavior dominated by $-C_6r^{-6}$. 
This common characteristic enabled a simple choice of characteristic length for 
classical and quantum systems, the van der Waals length, which was used for 
defining the scaling energy and size. 

The universal law, which relates scaled size and energy, has been found for 
homogeneous pentamers in their ground-state. 
For medium and weakly bound systems, it shows a non-linear non-logarithmic 
shape~\eqref{eq:fit} valid for scaled energies $X_E<10^6$, while after 
$X_E>10^3$ it approaches its asymptotic simple linear shape~\eqref{eq:YCL} in 
log-log scale. 
The law is applicable if the pair potential asymptotically follows as $-C_6r^{-6}$, while slower or faster decrease would produce a universal law with different log-log slope, as it can be deduced from the classical analysis. 
In the limit of the binding threshold, the scaled size of homogeneous pentamers
monotonously converge to the finite value $15$, below the tetramer size 
of $25$~\cite{Uni4}. Noticeably, this plateau is not present in the case of 
dimers and trimers which instead show a diverging size approaching the 
threshold for binding~\cite{RMPhalos,Unihalo,Efimov8}.

The universal size-energy line is also applicable to mixed systems which are 
homogeneous-like, i.e., if the distance between all the constituents is 
similar.  
If the mean square distance of the largest particle pair is few times larger 
than the shortest one, then mixed system could deviate above the line. 
In addition we find that if the cluster is spatially constrained, its size is 
reduced, so it can appear slightly below the line. 

Finally, we analyzed the relationship of pentamer, tetramer, and trimer 
energies of homogeneous systems, confirming the range of approximate validity of 
generalized Tjon lines and demonstrating the convergence of Lennard-Jones 
systems to the classical limit.


\section*{Acknowledgements}
	This work has been supported in part by the Croatian Science Foundation under the project number IP-2014-09-2452 and University of Split, Faculty of Science. 
	Partial financial support from   MCIN/AEI/10.13039/501100011033 (Spain)  
	grant No.  PID2020-113565GB-C21 is also acknowledged. J. B.  acknowledges 
	financial support from Secretaria d'Universitats i Recerca del Departament 
	d'Empresa i Coneixement de la Generalitat de Catalunya, co-funded by the 
	European Union Regional Development Fund within the ERDF Operational Program of 
	Catalunya (project QuantumCat, ref. 001-P-001644).
	
	The computational resources of the Isabella cluster at Zagreb University Computing Center (Srce), Croatian National Grid Infrastructure (CRO NGI) and server UniST-Phy at the University of Split were used.
	
	All figures were plotted by the command-line driven graphing utility Gnuplot, version 5.4, URL: \href{http://www.gnuplot.info}{www.gnuplot.info}. 
	
	This version of the article has been accepted for publication, after peer review but is not the Version of Record and does not reflect post-acceptance improvements, or any corrections. The Version of Record is available online at:\\ \href{http://dx.doi.org/10.1038/s41598-022-13630-2}{http://dx.doi.org/10.1038/s41598-022-13630-2}.

\section*{Author contributions}
    P.S. calculated ground-state properties and fitted the results. L.V.M. compared data with the generalized Tjon lines. All authors discussed the results and reviewed the manuscript. 

\section*{Additional information}
	\noindent\textbf{Competing interests}: The authors declare no competing interests.
	
	\noindent\textbf{Correspondence and requests for materials} should be addressed to P.S.
	
	\noindent\textbf{Data availability:} 
	The data that support the findings of this study are available within the article and its Supplementary Information \href{https://doi.org/10.1038/s41598-022-13630-2}{https://doi.org/10.1038/s41598-022-13630-2}.
	
\listoffigures
\listoftables

\end{document}